\documentclass[aps,prb,preprint,superscriptaddress]{revtex4}
\usepackage{graphicx}% Include figure files
\usepackage{dcolumn}% Align table columns on decimal point
\usepackage{bm}% bold math

\begin{document}

\title{Fermi surface evolution through a heavy fermion
superconductor-to-antiferromagnet transition: de Haas-van Alphen
effect in Cd-substituted CeCoIn$_5$}

\author{C. Capan}
%\email[]{Your e-mail address}
%\homepage[]{Your web page}
%\thanks{}
%\altaffiliation{}
\affiliation{Department of Physics and Astronomy, University of
California Irvine, Irvine, CA 92697-4575}

\author{Y-J. Jo}
%\email[]{Your e-mail address}
%\homepage[]{Your web page}
%\thanks{}
%\altaffiliation{}
\affiliation{National High Magnetic Field Laboratory, Florida State
University, Tallahassee, Florida 32310}

\author{L. Balicas}
%\email[]{Your e-mail address}
%\homepage[]{Your web page}
%\thanks{}
%\altaffiliation{}
\affiliation{National High Magnetic Field Laboratory, Florida State
University, Tallahassee, Florida 32310}

\author{R. G. Goodrich}
\altaffiliation[Present address: ]
                     {Department of Physics, George Washington University, Washington DC 20052}
\affiliation {Department of Physics and Astronomy, Louisiana State
University, Baton Rouge, Louisiana 70803}

\author{J. F. DiTusa}
\affiliation {Department of Physics and Astronomy, Louisiana State
University, Baton Rouge, Louisiana 70803}

\author{I. Vekhter}
\affiliation {Department of Physics and Astronomy, Louisiana State
University, Baton Rouge, Louisiana 70803}

\author{T. P. Murphy}
\affiliation{National High Magnetic Field Laboratory, Florida State
University, Tallahassee, Florida 32310}

\author{A. D. Bianchi}
\altaffiliation{Present address: Department de Physique, Universite
de Montreal, Montreal H3C 3J7 Canada}\affiliation{Department of
Physics and Astronomy, University of California Irvine, Irvine, CA
92697-4575}

\author{L.D. Pham}
\altaffiliation{Intel Corp.} \affiliation{Department of Physics and
Astronomy, University of California Irvine, Irvine, CA 92697-4575}

\author{J. Y. Cho}
\affiliation {Department of Chemistry, Louisiana State University,
Baton Rouge, Louisiana 70803}

\author{J. Y. Chan}
\affiliation {Department of Chemistry, Louisiana State University,
Baton Rouge, Louisiana 70803}

\author{D. P. Young}
\affiliation {Department of Physics and Astronomy, Louisiana State
University, Baton Rouge, Louisiana 70803}

\author{Z. Fisk}
\affiliation{Department of Physics and Astronomy, University of
California Irvine, Irvine, CA 92697-4575}

%Collaboration name if desired (requires use of superscriptaddress
%option in \documentclass). \noaffiliation is required (may also be
%used with the \author command).
%\collaboration can be followed by \email, \homepage, \thanks as well.
%\collaboration{}
%\noaffiliation

\date{\today}

\begin{abstract}
We report the results of de-Haas-van-Alphen (dHvA) measurements in Cd
doped CeCoIn$_5$ and LaCoIn$_5$. Cd doping is known to induce an
antiferromagnetic order in the heavy fermion superconductor
CeCoIn$_5$, whose effect can be reversed with applied pressure. We
find a slight but systematic change of the dHvA frequencies with Cd
doping in both compounds, reflecting the chemical potential shift due
to the addition of holes. The frequencies and effective masses are
close to those found in the nominally pure compounds with similar
changes apparent in the Ce and La compounds with Cd substitution. We
observe no abrupt changes to the Fermi surface in the high field
paramagnetic state for $x \sim x_c$ corresponding to the onset of
antiferromagnetic ordering at $H=0$ in CeCo(In$_{1-x}$Cd$_x$)$_5$. Our
results rule out $f-$electron localization as the mechanism for the
tuning of the ground state in CeCoIn$_5$ with Cd doping.
\end{abstract}

% insert suggested PACS numbers in braces on next line
% insert suggested keywords - APS authors don't need to do this
%\keywords{}

%\maketitle must follow title, authors, abstract, \pacs, and \keywords
\maketitle

\section{Introduction}
\indent A common thread in unconventional superconductivity is that
it emerges close to a quantum critical point (QCP). A QCP is the
point in a phase diagram where long-range order is suppressed to
zero temperature, $T=0$, by an external parameter other than $T$ so
that quantum, rather than thermal fluctuations drive the
transition\cite{QCP}. One way to rationalize the QCP in heavy
fermion metals is the phase diagram proposed by
Doniach\cite{Doniach} in which the ground state evolves from a local
moment antiferromagnet to a heavy fermion paramagnet as a function
of the tuning parameter $J g(\varepsilon_F)$, where $J$ is the
exchange coupling strength and $g(\varepsilon_F)$ the density of
states at the Fermi level. The QCP then corresponds to the point
where the Kondo energy scale $\sim exp(-\frac{1}{J
g(\varepsilon_F)})$ equals the RKKY scale $\sim (J g(\varepsilon_F)
^2)$. The quantum-critical spin-fluctuations associated with the
suppression of antiferromagnetic order are likely involved in the
pairing mechanism for unconventional superconductivity
\cite{Mathur98, Monthoux07}. This picture has been considered as an
explanation for a broad range of superconductors, including
high-$T_c$ cuprates\cite{Tallon20049}, heavy fermion
metals\cite{CeCu2Si2-1979,UBe13-1983,petrovic-2001-13,kuga:137004},
cobaltates\cite{Ihara20072119}, as well as the recently discovered
iron-pnictides\cite{2009PNAS}. However, there are also important
exceptions where unconventional superconductivity is observed and no
competing magnetic order is found, such as in the cases of
UBe$_{13}$\cite{UBe13-1983} and Sr$_2$RuO$_4$\cite{Maeno}.  There is
also the possibility of valence, rather than spin, fluctuation
mediated superconductivity as suggested for CeCu$_2$Si$_2$ under
pressure\cite{CeCu2Si2-1979}.\\

\indent Many studies have used doping as a tuning parameter between
superconducting and antiferromagnetic ground states in a broad range
of strongly correlated electron systems hosting a
QCP\cite{RevModPhys.81.1551,2008PNAS,Ihara20072119,2009PNAS}. The
heavy fermion metals in particular are very susceptible to chemical
substitution. In these compounds the Kondo coupling between a
lattice of local moments and the conduction band creates
quasi-particle excitations with large effective masses, and the
dopants disrupt the coherent Kondo coupling. Such studies have been
essential in assessing the percolative nature in which the coherence
in the Kondo lattice emerges -see for example the La-dilution study
of CeCoIn$_5$\cite{twofluid}- as well as its sensitivity to
disorder\cite{JPnature07}. But doping can also tune the ground state
by changing the carrier concentration as is remarkably illustrated
in the high-T$_c$ cuprates\cite{Tallon20049}. One important aspect
of doping the CuO$_2$ layers in the high-T$_c$ cuprates is the
apparent electron-hole symmetry: the phase diagrams are
qualitatively similar whether the carriers introduced are
electron-like or hole-like. One can then focus on a universal phase
diagram as a function of the carrier concentration, without having
to investigate the local effects associated with each particular
dopant. This symmetry is not found in CeCoIn$_5$ and so it is not
possible to define a universal phase diagram with doping as we
demonstrate below.\\

\indent CeCoIn$_5$ is a heavy fermion
superconductor\cite{petrovic-2001-13} where Cooper pairs are formed
out of a non-Fermi Liquid metallic state. The divergence observed in
the electronic specific heat, as well as the non-quadratic
$T$-dependence of the resistivity found even at very low
temperatures, suggest the presence of a QCP when superconductivity
is suppressed by a magnetic field\cite{Andrea-03,JP-03}. The nature
of the QCP has been the subject of much speculation, but it seems
likely to be an antiferromagnetic QCP. Hall effect measurements
under pressure have shown that the QCP is located not exactly at the
upper critical field $H_{c2}$ but at a slightly lower
field\cite{singh:057001}. Inelastic neutron
scattering\cite{stock:087001} and NMR
measurements\cite{young:036402}, on the other hand, have revealed
the presence of antiferromagnetic fluctuations within the
superconducting state. More recently, a field induced
antiferromagnetic order coupled to superconductivity has been
discovered close to $H_{c2}$ via neutron
scattering\cite{M.Kenzelmann09192008} and $\mu$SR
measurements\cite{PhysRevLett.103.237003} in pure CeCoIn$_5$.

\indent The ability to grow sizable, high quality, single crystals
enables detailed investigation of the effect of chemical doping in
this and other 115 compounds. While Sn-doping was found to suppress
$T_c$ without revealing any incipient magnetism\cite{bauer:047001},
Cd doping induces an antiferromagnetic ground state in CeCoIn$_5$.
The same behavior is also observed in the two other stoichiometric
Ce$\emph{M}$In$_5$ ($\emph{M}=$Rh, Ir)\cite{pham:056404} as well as
the bilayer Ce$_2\emph{M}$In$_8$($\emph{M}=$Co, Rh,
Ir)\cite{adriano-2009} with Cd doping. Because Sn and Cd are
neighbors to In in the periodic table, Sn and Cd substitutions for
In result in electron and hole doping, respectively. The effect of
Cd is quite unusual in the sense that it takes a very small density
of Cd to induce the paramagnetic to antiferromagnetic (AFM) ground
state transformation which can be reversed with the application of
pressure\cite{pham:056404}. How Cd induces long range AFM order with
a large ordered magnetic moment\cite{urbano:146402} (0.7$\mu_B$ per
Ce) in CeCoIn$_5$ remains an open question.

\indent One possible mechanism is the formation of antiferromagnetic
droplets at the Cd sites, as was inferred from NMR
measurements\cite{urbano:146402}. Long range AFM order occurs once
the density of such droplets reaches the percolation threshold.
However, the density of Cd necessary to induce ordering is well
below the percolation threshold. Thus the ordering at such a small
Cd concentration requires very long correlation length and
correspondingly large size of the ordered droplet around each
dopant. Thus the ordering at such a small Cd concentration requires
interactions with a longer range. Since the ordered moments are
likely local moments on the Ce sites, one way to account for the
reversibility of Cd doping with pressure is to speculate that the
change in carrier density and disorder caused by Cd substitution
localizes the $f$-electrons of nearby Ce atoms. Application of
pressure to metals with localized f-orbitals tends to increase the
hybridization with the conduction band and delocalize f-electrons.

\indent Alternatively, the AFM state is due to a Fermi Surface (FS)
instability, which is the well-known explanation in the case of
elemental Cr. Recent neutron scattering results in Cd doped
CeCoIn$_5$\cite{nicklas:052401} have demonstrated that the AFM
ordering has a wavevector, $Q$ of (1/2,1/2,1/2) suggesting that if
the AFM is nesting-driven, the nesting wavevector is commensurate
with the lattice. This is a plausible, but unusual, situation that
occurs, for example, in Mn doped Cr\cite{fawcett}. It is possible
that in the situation intermediate between local and itinerant the
magnetic ordering is driven by the local, unscreened component of
the spin, and the improved near-nesting with Cd doping lowers the
energy cost of opening the gap for itinerant electrons at the
magnetic Brillouin Zone boundary, enabling the long-range order to
appear.

\indent Perhaps the most surprising feature of this ordered state is
that the ordering wavevector\cite{nicklas:052401} coincides with the
wavevector at which an inelastic neutron scattering
resonance\cite{stock:087001} is observed in the superconducting
state of pure CeCoIn$_5$. The origin of this resonance has been
attributed to AFM magnons\cite{chubukov:147004} and we suspect the
coincidence is not accidental. A similar AFM state can be induced
with Rh substitution for Co in CeCoIn$_5$ for Rh concentrations
greater than $\sim 25\%$ \cite{PhysRevB.65.014506}. Here, the AFM
wavevector is identical to that found for Cd doped CeCoIn$_5$, which
coexists with superconductivity for Rh concentrations of less than
$\sim 60\%$. For larger Rh concentrations the AFM wavevector becomes
(1/2,1/2,$\sim 0.3$) and superconductivity is suppressed. The change
in wavevector and the loss of superconductivity at Rh concentrations
above $60\%$ suggest that the AFM state in Rh (for $x_{Rh}<0.6$) and
Cd doped CeCoIn$_5$ has a different character from the local moment
antiferromagnetism found in CeRhIn$_5$. In order to understand more
fully the superconducting state in CeCoIn$_5$ and the relevance of
the nearby AFM QCP, the character of the AFM state will require
further investigations.

\indent In this paper, we report on one such investigation by
specifically exploring the evolution of the FS of both CeCoIn$_5$
and LaCoIn$_5$ as a function of Cd doping via de Haas van Alphen
(dHvA) oscillations. We observe that the changes to the FS with Cd
substitution are consistent with the addition of holes and that the
FS varies with Cd substitution at a similar rate in both compounds.
Our results thus rule out $f$-electron localization as a possible
route towards AFM order. If this were the case it would lead to a
significantly different evolution of the Fermi surfaces of
CeCo(In$_{1-x}$Cd$_x$)$_5$ and the non-magnetic La-analog. This
paper is organized as follows: We first present the experimental
details in section II followed by an introduction to the phase
diagram in section III. Section IV focuses on the evolution of the
Fermi Surface in Cd doped CeCoIn$_5$ in comparison to Cd doped
LaCoIn$_5$ while section V presents the effect of Cd on the
cyclotron effective mass and mean free path. We summarize our
findings in section VI and discuss the possible mechanism(s) for AFM
order in Cd doped CeCoIn$_5$ that is (are) consistent with our data.

\section{Experimental Details}

\begin{figure}
\resizebox{!}{0.6\textwidth}{\includegraphics{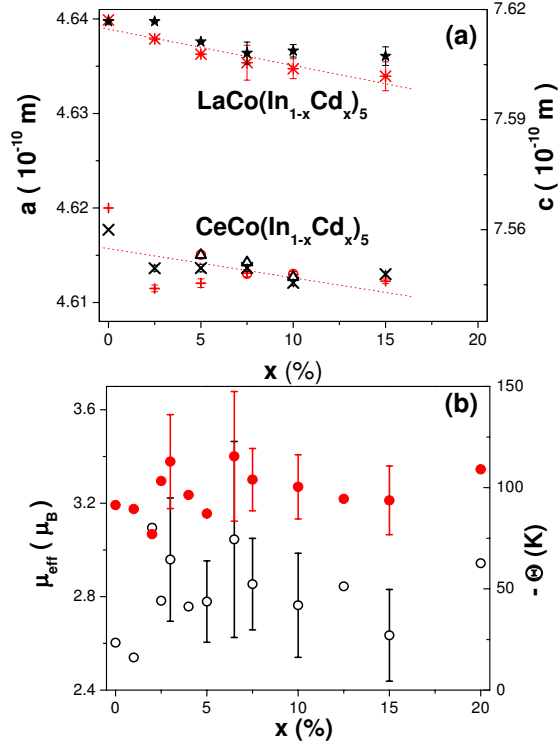}}
\caption{\label{fig:xrays} (Color online) (a) Lattice parameters $a$
and $c$ along [100] and [001] respectively vs nominal Cd
concentration $x$ in LaCo(In$_{1-x}$Cd$_x$)$_5$ ($\ast: a$, $\star:
c$) and CeCo(In$_{1-x}$Cd$_x$)$_5$ ($+: a$, $\times: c$) obtained
from powder X-ray diffraction. Single crystal X-ray diffraction
results are also shown for $x=5\%, 7.5\%$ and $10\%$ in
CeCo(In$_{1-x}$Cd$_x$)$_5$ ($\circ: a$, $\bigtriangleup: c$). The
error bars are smaller than the symbol size where they are not
shown. The dotted lines are linear fits to $a$ vs $x$. (b)
Curie-Weiss moment $\mu_{eff}$ ($\circ$) and Curie-Weiss Temperature
$\Theta$ ($\bullet$) vs nominal Cd concentration $x$ in
CeCo(In$_{1-x}$Cd$_x$)$_5$. The Curie-Weiss parameters were obtained
from fits to the susceptibility in the $T-$range of $100-400$~K,
measured with $H=0.1$~T applied perpendicular to [001]. For
concentrations in which more than one sample was measured average
values are shown with error bars corresponding to the standard
deviations.}
\end{figure}

\indent The single crystals of CeCo(In$_{1-x}$Cd$_x$)$_5$ and
LaCo(In$_{1-x}$Cd$_x$)$_5$ used in our experiments are grown from In
flux in a ratio of Ln:Co:In:Cd ($1:1:20(1-x):20x$) from high purity
starting materials, as described elsewhere\cite{petrovic-2001-13}.
The lattice parameters were determined by using both powder and
single crystal X-ray diffraction and are shown in
figure~\ref{fig:xrays}a.  Si was used as a standard in the Rietveld
refinement of the powder X-ray diffraction patterns.  We have
determined the Cd concentration via Energy Dispersive X-ray analysis
(EDS) resulting in values comparable to those published by other
groups\cite{tokiwa:037001}.  These measurements indicate that only a
fraction of the Cd ($\sim 30\%$) effectively substitutes for In, as
we found $x=1.6$, 1.9 and $2.3\%$ for nominal concentrations of
$x=2.5$, 5 and $7.5\%$ in CeCo(In$_{1-x}$Cd$_{x}$)$_5$. Similarly,
we obtained $x=1.3$, 1.6, 2.2 and $2.3\%$ for nominal concentrations
of $x=2.5$, 5 , 7.5 and $10\%$ in LaCo(In$_{1-x}$Cd$_{x}$)$_5$. For
ease of comparison with previous work we quote the nominal
concentrations throughout the paper.

\indent The susceptibility of single crystals of
CeCo(In$_{1-x}$Cd$_x$)$_5$ was measured at $H=0.1$~T applied
perpendicular to [001] for temperatures ranging from 1.8 to 400 K
using a commercial vibrating sample Superconducting Quantum
Interference Device (SQUID) magnetometer. The Curie-Weiss parameters
were obtained from fits to susceptibility in the $T-$range of
$100-400$~K and are shown in figure~\ref{fig:xrays}b. For the
concentrations for which more than one sample was measured, the
average values are shown and the error bars correspond to the
standard deviation. The resistivity was measured from 1.8 to 300~K
at $H=0$ with a current of 1~mA applied along [100] in single
crystals of CeCo(In$_{1-x}$Cd$_x$)$_5$ for $x=5, 10$ and $15\%$. The
crystals were In free and the Pt wires were attached using silver
epoxy.

\indent The evolution of the FS of the same crystals used for the
single crystal X-ray diffraction measurements was investigated via
the Fast Fourier Transform (FFT) analysis of the dHvA oscillations
measured using a torque magnetometer\cite{Shoenberg}.  Single
crystals were mounted on a Cu-Be cantilever, inside either a
$^{3}$He cryostat or a dilution refrigerator equipped with a
rotator.  The torque signal was measured at the National High
Magnetic Field Laboratory, using a capacitance bridge in magnetic
fields of up to 35~T, and for temperatures down to 0.3~K and 0.05~K
for the LaCo(In$_{1-x}$Cd$_x$)$_5$ and the
CeCo(In$_{1-x}$Cd$_x$)$_5$ crystals, respectively. For the FFT
analysis it is assumed that $H \approx B$ without demagnetizing
factor correction, since we estimate the magnetization of
CeCo(In$_{1-x}$Cd$_x$)$_5$ to be $\sim 0.4\%$ of the applied field.
Indeed, the in-plane susceptibility (measured down to 1.8~K) is
extrapolated with a power law fit to $\chi_{\perp}=0.0143$~emu/mol
at $T=50$~mK in $5\%$Cd doped CeCoIn$_5$, which corresponds to
$\chi_{\parallel}=0.0286$~emu/mol for $H \parallel [001]$ (with a
magnetic anisotropy of 2) and to a volume magnetization of $4 \pi M
= 1260$~G at 35~T.

\indent Fig.~\ref{fig:xrays}a shows the change in lattice parameters
as a function of the nominal Cd concentration in
LaCo(In$_{1-x}$Cd$_x$)$_5$ and CeCo(In$_{1-x}$Cd$_x$)$_5$ where we
observe that the main effect of Cd substitution is to produce a volume
contraction in both compounds. This is as expected since Cd atoms are
smaller than In atoms. The volume contraction rate is similar in both
the Ce and the La compounds and is in quantitative agreement with the
contraction inferred from a local structure, extended X-ray absorption
fine structure (EXAFS), investigations\cite{booth:144519} for the
same nominal concentrations. EXAFS results also indicate that Cd, as
well as Sn, preferentially substituted for In on the in-plane, In(1),
site\cite{booth:144519}. The close agreement between the powder and
the single crystal X-ray lattice parameters (see
fig.~\ref{fig:xrays}a) suggest that the variation of the Cd
concentration within a batch is small: we estimate a difference of
$\Delta x \leq 2\%$ between single crystal and average (powder)
nominal concentrations.

\indent The similar suppression of the unit cell volume in both the
Ce- and the La-compounds apparent in Fig.~\ref{fig:xrays}a indicates
that Cd doping has no significant effect on the valence of Ce at
room $T$.  If the addition of Cd were to change the valence of Ce,
the size of the Ce ions, and consequently the lattice parameters of
CeCo(In$_{1-x}$Cd$_x$)$_5$, would have a rather distinct doping
dependence in comparison to their La-analogs. This is also supported
by the lack of a systematic variation of either the effective Curie
moment or the Weiss temperature with Cd substitution
(Fig.~\ref{fig:xrays}b) as determined from the magnetic
susceptibility. However, these data do not preclude a possible
valence fluctuation scenario at low $T$.

\section{The Phase Diagram}

\begin{figure}
\resizebox{!}{0.6\textwidth}{\includegraphics{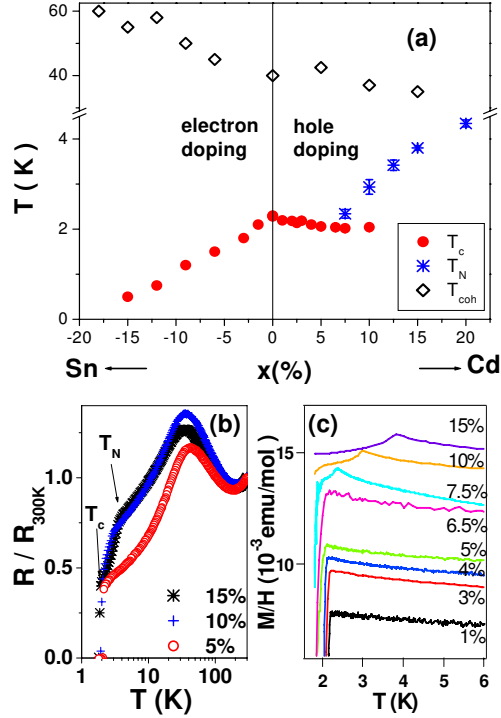}}
\caption{\label{fig:phasediagram} (Color online) Phase diagram. (a)
Temperature, $T$ vs. doping, $x$ phase diagram of
CeCo(In$_{1-x}$M$_x$)$_5$, M=Sn,Cd. The Sn and Cd concentrations are
nominal. The superconducting ($T_c, \bullet$) and the
antiferromagnetic ($T_N, \ast$) transition temperatures for Cd doped
samples are obtained from magnetic susceptibility. The Kondo
coherence temperature ($T_{coh}, \diamond$) corresponds to the
resistivity maximum.  $T_{coh}$ and $T_c$ for Sn doped samples are
from ref.~\onlinecite{bauer:245109}. (b) Normalized resistance,
$R/R_{300 K}$ vs. $T$ from 1.8 to 300~K on a semi-log plot at $H=0$
for single crystals of CeCo(In$_{1-x}$Cd$_x$)$_5$ with $x=5, 10$ and
$15\%$. (c) Magnetic susceptibility, $M/H$ vs. $T$ at $H=0.1$~T in
CeCo(In$_{1-x}$Cd$_x$)$_5$ in the range $1.8-6$~K showing the
superconducting and antiferromagnetic transitions. Nominal Cd
concentrations are indicated in the figure. The $x=5, 10\%$ and
$15\%$ data have been shifted by 2, 5 and $6\times 10^{-3}$ emu/mol
vertically for clarity.}
\end{figure}

\indent The doping dependent phase diagram of
CeCo(In$_{1-x}$Cd$_x$)$_5$ is shown in Fig.~\ref{fig:phasediagram}a
together with the resistivity and the magnetic susceptibility in
Fig.~\ref{fig:phasediagram}b and \ref{fig:phasediagram}c. The
superconducting critical temperature $T_c$ and the Neel temperature
$T_N$ are determined from the sharp drop and the peak in the
magnetic susceptibility, respectively (see
Fig.~\ref{fig:phasediagram}c). These are consistent with the
transitions seen in the resistivity at $H=0$ (see
Fig.~\ref{fig:phasediagram}b). Fig.~\ref{fig:phasediagram}a includes
the Sn-doping phase diagram\cite{bauer:245109} for comparison.
Superconductivity is suppressed with both Sn and Cd doping as a
result of pair-breaking via impurity scattering, although this
suppression appears to be stronger with Sn than for Cd dopants.
Antiferromagnetic order sets-in for $x\geq 7.5\%$ for Cd doping
only, emphasizing the electron-hole asymmetry in the doping
phase-diagram of CeCoIn$_5$.

\indent The overall phase diagram as a function of Cd doping
obtained by us is consistent with a previous
report\cite{pham:056404}, and in particular with a finite range of
coexistence for both the superconducting and AFM phases.  While the
samples with $x=7.5\%$ systematically show both superconducting and
AFM transitions, traces of superconductivity are also observed for
$x=10\%$, as seen in Fig.~\ref{fig:phasediagram}c, although this is
highly sample dependent. The superconducting transition is also
observed in the $H=0$ resistivity (see Fig.~\ref{fig:phasediagram}b)
for our $x=10$ and $15\%$ crystals. This suggests that the doping
may be somewhat inhomogeneous within a given single crystal.
Nevertheless, a microscopic coexistence of both orders has been
claimed based on neutron scattering and NMR
measurements\cite{nicklas:052401, urbano:146402}. The evolution of
transition temperatures $T_c$ and $T_N$ with Cd doping remains quite
systematic (see Fig.~\ref{fig:phasediagram}a) with only a small
variation observed within a given batch. Superconductivity
coexisting with a commensurate AFM order appears to be a generic
feature of doped CeCoIn$_5$ since it was also observed with Rh
substitution\cite{PhysRevB.65.014506,jeffries} for $x_{Rh}<0.6$.

\indent The Kondo coherence temperature $T_{\text{coh}}$ in
CeCo(In$_{1-x}$Cd$_x$)$_5$, as determined from the maximum in the
resistivity as a function of temperature (see
Fig.~\ref{fig:phasediagram}b), is displayed in
Fig.~\ref{fig:phasediagram}a.  The Cd doping tends to suppress
$T_{\text{coh}}$, a trend which is the opposite to the effect of Sn
doping\cite{bauer:245109} which is included in
Fig.~\ref{fig:phasediagram}a for comparison. Since $T_{\text{coh}}$
increases with pressure\cite{0953-8984-13-44-104} one way of
rationalizing the evolution of the $T_{\text{coh}}$ with Sn and Cd
doping is in terms of the lattice volume change. However, the
enhancement of $T_{\text{coh}}$ with Sn doping is not simply a
chemical pressure effect since Sn has no detectable effect on the
lattice volume\cite{bauer:245109}. Nor is the suppression of
$T_{\text{coh}}$ specific to Cd: a recent investigation on
rare-earth substitution has shown that $T_{\text{coh}}$ is
systematically suppressed as the Ce lattice is diluted, regardless
of the magnetic or electronic nature of the rare-earth dopants
\cite{JPnature07}. This fact, taken alone, may seem to suggest that
the small suppression of $T_{\text{coh}}$ with Cd doping is
effectively a dilution effect as Cd localizes the f-electrons on a
small number of neighboring Ce ions. As we see below, this is not
supported by our measurements. Moreover, there is an important
difference in that both Yb\cite{capan2009} and Cd\cite{pham:056404}
act as hole dopants but only Cd stabilizes the AFM state in
CeCoIn$_5$. The opposing effects of Sn and Cd instead suggest that
the changes to $T_{\text{coh}}$ are a consequence of the shift in
the chemical potential corresponding to electron or hole doping.

\indent Tuning the ground state with Cd does not appear to conform
to the Doniach phase diagram of competing RKKY and Kondo
scales\cite{Doniach} since no systematic change is observed in
either the Curie-Weiss temperature (see Fig.~\ref{fig:xrays}b), a
measure of the RKKY interaction strength, nor the single-ion Kondo
scale, determined from the magnetic entropy of a series of $5\%$ Ce
doped LaCo(In,Cd)$_5$ crystals (not shown), with Cd substitution.
NMR measurements\cite{urbano:146402} also indicate an absence of
change to the low energy spin fluctuation spectrum with Cd
substitution in CeCoIn$_5$ in the paramagnetic state.  Thus, the
natural question is whether the Cd-induced antiferromagnetism is,
instead, due to a Fermi surface instability, a possibility we
investigate via the dHvA measurements presented below.

\section{The Fermi Surface}

\begin{figure}
\resizebox{!}{0.5\textwidth}{\includegraphics{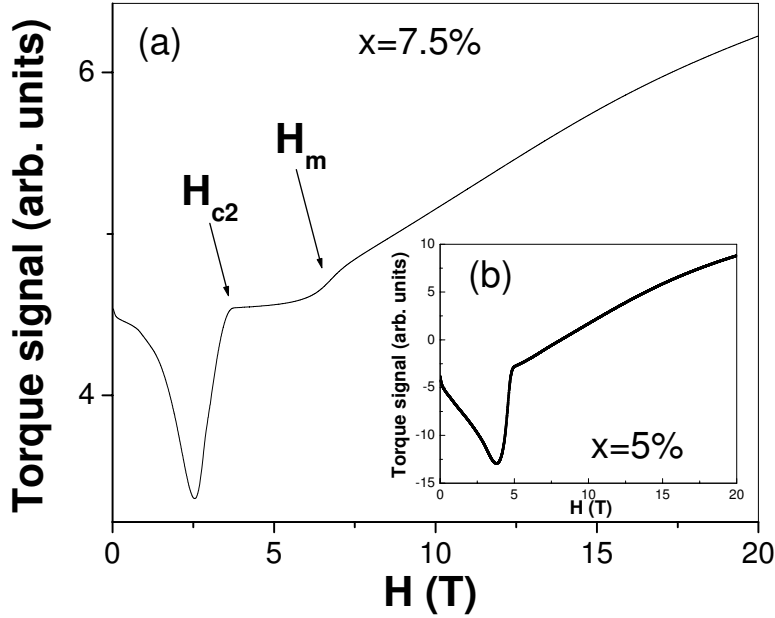}}
\caption{\label{fig:Torque} Torque signal. Torque signal vs magnetic
field in CeCo(In$_{1-x}$Cd$_x$)$_5$ at $T=0.05$ K for $x=7.5\%$ (a)
and $5\%$ (b). The torque signal is proportional to
magnetization\cite{Shoenberg}. The magnetic field is oriented at
$8^{o}$ from [001]. H$_{c2}$ and H$_m$ correspond to the
superconducting upper critical field and the metamagnetic transition
respectively.}
\end{figure}

\begin{figure}
\resizebox{!}{0.5\textwidth}{\includegraphics{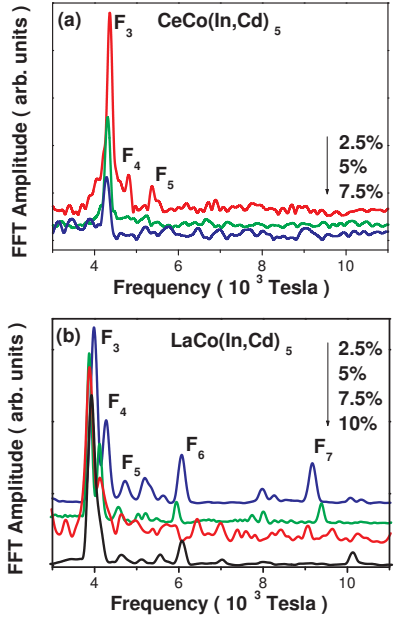}}
\caption{\label{fig:FFTspectra} (Color online) de Haas-van Alphen
Spectrum. Fast Fourier Transform (FFT) vs Frequency in: (a)
CeCo(In$_{1-x}$Cd$_x$)$_5$ and (b) LaCo(In$_{1-x}$Cd$_x$)$_5$. The
nominal concentrations of $x= 2.5\%, 5\%, 7.5\%$ and $10\%$ are
indicated. The spectra shown for each concentration is taken for
magnetic field oriented at an angle $\Theta \leq 15^o$ from [001],
at $T=0.05$~K and $T=0.3$~K for CeCo(In$_{1-x}$Cd$_x$)$_5$ and
LaCo(In$_{1-x}$Cd$_x$)$_5$ respectively. The peaks $\emph{F}_3,
\emph{F}_4, \emph{F}_5$ ($\emph{F}_6,\emph{F}_7$) correspond to the
electron (hole) sheets of the Fermi Surface.}
\end{figure}

\indent To investigate the changes to the Fermi surface with Cd
substitution that coincide with the variations noted above, we have
systematically measured the dHvA oscillations as a function of $x$,
$T$, and the magnitude and direction of $H$.  Figure~\ref{fig:Torque}
shows the torque signal (in arbitrary units) as a function of magnetic
field between 0 and 20~T oriented at $8^{o}$ from [001] at 0.05 K in
CeCo(In$_{1-x}$Cd$_x$)$_5$ for $x=5$ and $7.5\%$. At these low fields
dHvA oscillations are not yet apparent. While both samples exhibit a
pronounced dip in the vortex state, there is a distinct metamagnetic
anomaly at $H_m \simeq 7$~T $>H_{c2}$ for the $x=7.5\%$ sample,
corresponding to the transition from the antiferromagnetic to the
paramagnetic state. A maximum in the transverse MR of an $x=0.1$
sample is observed at around the same field (not shown). Thus, it
appears that for fields large enough for dHvA oscillations to be
detected the samples are in the high field paramagnetic state, rather
than in the zero field AFM phase. This restriction precludes the
observation of a Fermi surface reconstruction in the magnetic
Brillouin zone of the AFM state. Despite this limitation we can learn
much about the Fermi surface and the mechanism for AFM in
CeCo(In$_{1-x}$Cd$_x$)$_5$ from our dHvA measurements.

\indent Figure~\ref{fig:FFTspectra} shows the Fast Fourier Transform
(FFT) of the torque signal (after background subtraction) as a
function of frequency in CeCo(In$_{1-x}$Cd$_x$)$_5$ and
LaCo(In$_{1-x}$Cd$_x$)$_5$ for all Cd concentrations measured. The
FFT is calculated on the same field range 25-35~T for all samples
and all orientations. No dHvA oscillations are resolved in
CeCo(In$_{1-x}$Cd$_x$)$_5$ for $x \geq 10\%$ for fields up to 45~T
and for temperatures down to 0.05~K. The peaks in the FFT spectra
shown in Fig.~\ref{fig:FFTspectra} correspond to the branches of the
electron and hole sheets of the Fermi surface that have been
previously identified\cite{settai2001,donavan2001}. The labeling of
these branches is identical to Ref.~\onlinecite{donavan2001}.
Overall, similar branches are observed in both Ce and La analogs,
with systematically larger frequencies in the Ce compounds as
compared to their La counterparts. This is also the case for pure
CeCoIn$_5$ and is due to the itinerant nature of the $4f$ electrons
in the sense that they are hybridized with the conduction
bands\cite{PhysRevLett.93.186405,shishido2002}. In contrast, the FS
of the antiferromagnetic compound CeRhIn$_5$ is known to be very
close to its non-magnetic analog LaRhIn$_5$, suggesting localized
$f-$electrons\cite{PhysRevLett.93.186405,shishido2002}.  The
incommensurate AFM order with a large moment $\mu \sim 0.8\pm 0.1
\mu_B$, in CeRhIn$_5$\cite{PhysRevB.62.R14621} is therefore a local
moment ordering similar to the incommensurate local moment magnetism
found in other rare-earths metals\cite{jensen}.

\begin{figure}
\resizebox{!}{0.6\textwidth}{\includegraphics{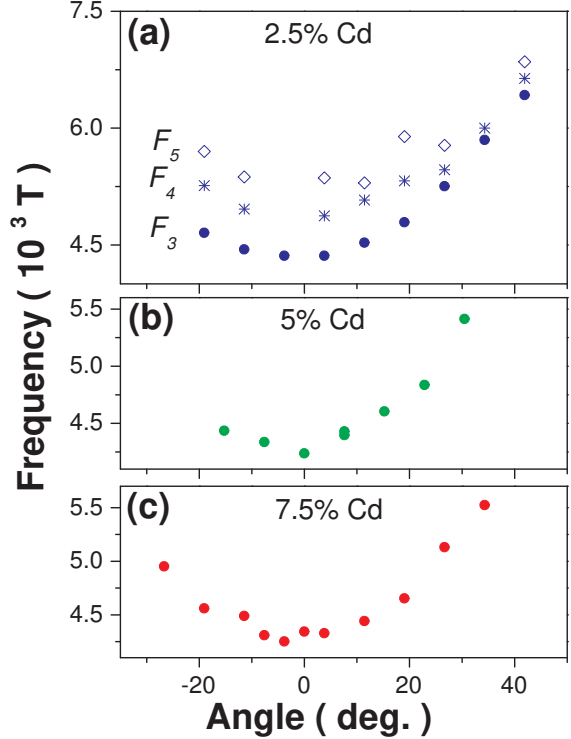}}
\caption{\label{fig:FreqvsAngle} (Color online) dHvA Frequency vs.
Angle in CeCo(In$_{1-x}$Cd$_x$)$_5$. Angular dependence shown for a)
$x=2.5\%$, b) $x= 5\%$ and c) $x=7.5\%$. Only the $\emph{F}_3$
branch is resolved in the $x=5\%$ and $7.5\%$ samples. Zero angle
corresponds to $H \parallel [001]$.}
\end{figure}

\indent The angular dependence of several dHvA branches is shown for
CeCo(In$_{1-x}$Cd$_x$)$_5$ in figure~\ref{fig:FreqvsAngle}. No
significant change is observed with Cd doping. It was previously
established that for nominally pure CeCoIn$_5$ the angular dependence
for most branches is well described by a $1 / \cos{\Theta}$ dependence
indicative of a quasi-two dimensional Fermi
surface\cite{donavan2001,settai2001} and this continues to be true for
the Cd doped samples.

\begin{figure}
\resizebox{!}{0.6\textwidth}{\includegraphics{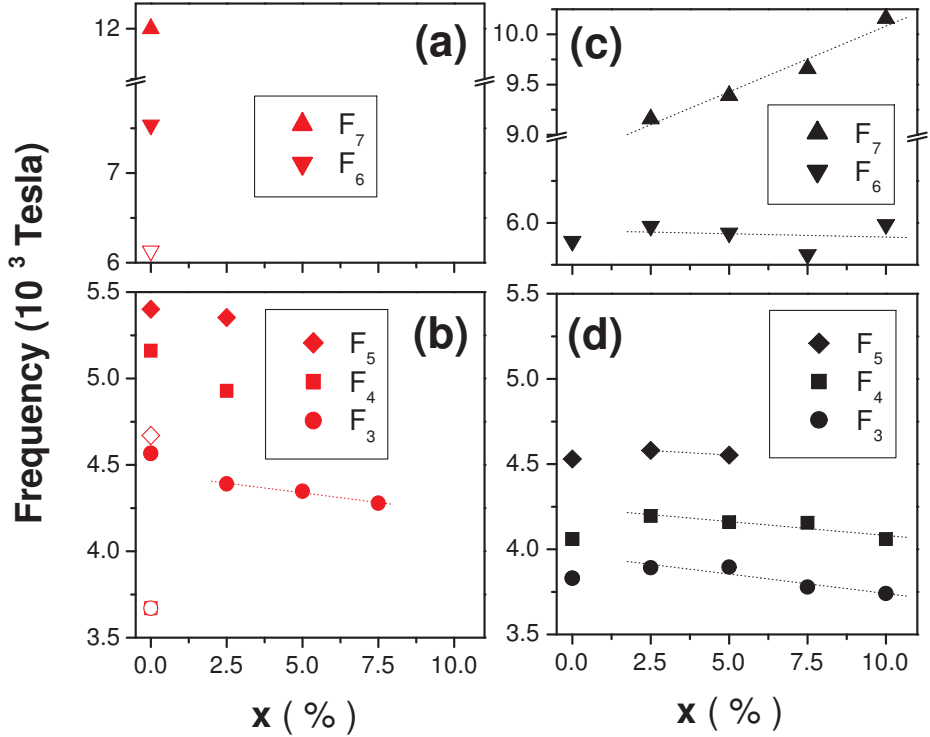}}
\caption{\label{fig:FreqvsCdAll}(Color online) Cd concentration
dependence of dHvA spectra. dHvA Frequency vs.  nominal Cd
concentration in CeCo(In$_{1-x}$Cd$_x$)$_5$ (panels $a,b$) and
LaCo(In$_{1-x}$Cd$_x$)$_5$ (panels $c,d$). The upper (lower) panels
$a,c$ ($b,d$) correspond to the various frequencies of the hole
(electron) sheet of the Fermi Surface for $H \parallel [001]$, as
indicated. The frequencies for pure CeCoIn$_5$, LaCoIn$_5$ and
CeRhIn$_5$ (open symbols) are from previously published dHvA
data\cite{PhysRevLett.93.186405,donavan2001,settai2001,shishido2002,roy2009}.
The dotted lines are linear fits to the data.}
\end{figure}

\indent The evolution of the dHvA frequencies with $x$ for
$H\parallel [001]$ is shown in figure~\ref{fig:FreqvsCdAll} for
CeCo(In$_{1-x}$Cd$_x$)$_5$ (panels $a,b$) and
LaCo(In$_{1-x}$Cd$_x$)$_5$ (panels $c,d$). The values reported in
the figure correspond to the minimum of the frequency vs angle
curves obtained via quadratic fits to the data in
fig.~\ref{fig:FreqvsAngle}. The LaCo(In$_{1-x}$Cd$_x$)$_5$
frequencies compare well with those previously reported for pure
LaCoIn$_5$\cite{roy2009}, also shown in fig.~\ref{fig:FreqvsCdAll}.
We have included data for nominally pure CeCoIn$_5$ (full symbols)
and CeRhIn$_5$ (open symbols) for the same branches and orientation
($H\parallel [001]$) taken from the
literature\cite{PhysRevLett.93.186405,donavan2001,settai2001,shishido2002}
in fig.~\ref{fig:FreqvsCdAll}. This comparison of the dHvA
frequencies for these two systems demonstrates that the substantial
differences, which were independently observed by two
groups\cite{shishido2002,PhysRevLett.93.186405}, are real and beyond
experimental uncertainty. The conclusion is that in CeCoIn$_5$ the
dHvA frequencies correspond to a "large" Fermi Surface which
includes a contribution from itinerant $f-$electrons, whereas in
CeRhIn$_5$ the dHvA frequencies correspond to a "small" Fermi
Surface expected in the case of well localized f-electrons..

\indent In Fig.~\ref{fig:FreqvsCdAll} we observe that the dHvA
frequencies of CeCo(In$_{1-x}$Cd$_x$)$_5$ remain very close to those
of the pure CeCoIn$_5$ with no abrupt change to $F_3$ at the
critical concentration, $x_c=7.5\%$. In fact, the rate at which the
$F_3$ frequency (electron orbit) is suppressed with Cd is very
similar, within the limits of our measurements, in the Ce and the La
analogs as emphasized by the linear fits in
Fig.~\ref{fig:FreqvsCdAll}. These observations tend to rule out any
change in the Ce valence at the lowest temperatures, as was inferred
at room $T$ from the lattice parameter evolution. Indeed, if the Ce
valence, or simply the hybridization of the $4f$ electron with the
conduction band through Kondo effect, changed with Cd in such a way
that the $4f$ electron was more localized, the effective number of
carriers introduced on the Fermi surface by each Cd would \emph{not}
be one hole as is expected to be for the case of
LaCo(In$_{1-x}$Cd$_x$)$_5$. One would then expect the $F_3$ branch
to decrease at a faster rate in CeCo(In$_{1-x}$Cd$_x$)$_5$ when
compared to its La-analog. This suggests that the FS of the
$x_c=7.5\%$ sample is significantly larger than that of the
antiferromagnetic counterpart CeRhIn$_5$ and rules out the
localization of $f-$electrons at the onset of antiferromagnetism at
$H=0$. The nucleation of a local moment AFM state at $x_c$ with a
large ordered moment, $\mu=0.7 \mu_B$ per Ce, would require a
substantial density of localized Ce $f$-electrons. This, in turn,
would requires a substantial change in the dHvA frequencies, of
order the difference between CeCoIn$_5$ and CeRhIn$_5$, with Cd
doping.  The similar evolution of the Fermi Surface in
CeCo(In$_{1-x}$Cd$_x$)$_5$ and its La analogs suggests that the
effect of Cd is primarily a rigid band shift (or equivalently a
shift in the chemical potential) due to the additional hole in
\emph{both} systems, without significant change in the Ce valence or
the Kondo hybridization of the $4f$ electron.

\indent This is perhaps the most important finding of our
investigation; the $f-$electrons in CeCo(In$_{1-x}$Cd$_x$)$_5$
remain itinerant with Cd doping. This is in stark contrast to the
naive expectation based upon the similarities in critical
temperatures and sizes of magnetic moments, $\mu$, that the
mechanism for magnetism in CeCo(In$_{1-x}$Cd$_x$)$_5$ at $x>x_c$ and
CeRhIn$_5$ are identical. In CeRhIn$_5$ the magnetism has been shown
to be due to RKKY coupling of the well localized Ce $f$-electron
magnetic moments, with an incommensurate
wavevector\cite{PhysRevB.62.R14621}. In CeCo(In$_{1-x}$Cd$_x$)$_5$
the antiferromagnetism is commensurate\cite{nicklas:052401} and the
large Fermi surface we observe is a further indication that the
mechanism driving the magnetism may be quite different from
CeRhIn$_5$: the magnetic order involves SDW-like rearrangement of
the Fermi surface, rather than a dramatic change in volume as in the
f-electron localization scenario. That the Ce $f$-electrons remain
well hybridized with the conducting electrons is indicated by the
large coherence temperatures seen in Fig.~\ref{fig:phasediagram} and
the insensitivity of the dHvA frequencies to Cd doping. Thus, we
reach a conclusion similar to that of a recent investigation of
Rh-doped CeCoIn$_5$ which indicated no change of the $F_3$ frequency
with Rh doping through the Rh concentration, $x\sim25\%$, for which
a commensurate AFM order sets in.  The insensitivity of this dHvA
frequency to Rh doping also implies a "large" Fermi surface for Rh
concentrations where superconductivity coexists with the
commensurate AFM order\cite{goh:056402}.

\indent The second important finding is that the evolution of dHvA
frequencies with Cd doping is opposite for the electron and the hole
Fermi Surface sheets in LaCo(In$_{1-x}$Cd$_x$)$_5$
(Fig.~\ref{fig:FreqvsCdAll}). In Fig.~\ref{fig:FreqvsCdAll}c and d
we observe a systematic variation with $x$ in some of the dHvA
frequencies: the frequency of the electron orbits $F_3, F_4$
decrease, while that of the hole orbit, $F_7$, increases with
increasing $x$. This suggests that the electron Fermi surface
($F_3,F_4$) shrinks for increasing Cd concentration, while the hole
Fermi surface ($F_7$) expands. Note that the $F_6$ orbit, which
derives from the same hole FS, is relatively constant suggesting the
expansion of the hole sheet is not uniform. Overall, the FS
evolution in LaCo(In$_{1-x}$Cd$_x$)$_5$ can be simply understood as
a chemical potential shift: Cd effectively is a hole dopant since Cd
has one electron less than In.

\indent In the case of CeCo(In$_{1-x}$Cd$_x$)$_5$ we do not observe
the hole Fermi surface (up to 35~T for $x=2.5\%$ and $5\%$, up to
45~T for $x=7.5\%$); the oscillations from the hole orbits are
likely suppressed due to the disorder scattering introduced with Cd
impurities. This smearing may be stronger for hole than for electron
orbits as a consequence of their larger effective
masses\cite{settai2001, donavan2001}. The electron Fermi surface in
CeCo(In$_{1-x}$Cd$_x$)$_5$ also shrinks ($F_3$ decreases with $x$),
and we can safely interpret this as the effect of hole doping in
analogy with LaCo(In$_{1-x}$Cd$_x$)$_5$. Note that the reduction of
the volume of the electron FS sheet corresponding to the decrease of
$F_3$ is very modest and accounts for only $\sim 1/30$ of the hole
introduced by Cd (see Appendix). Therefore it is likely that the
added hole is mainly distributed over the parts of the Fermi surface
which we do not observe in CeCo(In$_{1-x}$Cd$_x$)$_5$. Since the
effective masses\cite{settai2001} and the hybridization
gap\cite{burch:054523} are known to be anisotropic, we cannot
exclude that the chemical shift due to Cd leads to a more dramatic
volume change on the hole sheet of the FS, and that this change
creates nesting conditions. Note that the suppression of the $F_4$
and $F_5$ electron orbits with Cd in both the Ce and La compounds
may indicate a more cylindrical (less corrugated along the $c$-axis)
electron sheet. Perhaps such small changes in corrugation also lead
to improved nesting along the c-axis. Since the wave vector remains
commensurate in the plane (1/2, 1/2), and it is only the c-axis
component that becomes commensurate in Cd doped CeCoIn$_5$, as
compared to pure CeRhIn$_5$\cite{PhysRevB.62.R14621}, small changes
in the Fermi surface may indeed cause this lock onto
commensurability.

\section{The Effective Mass and the Mean Free Path}

\begin{figure}
\resizebox{!}{0.5\textwidth}{\includegraphics{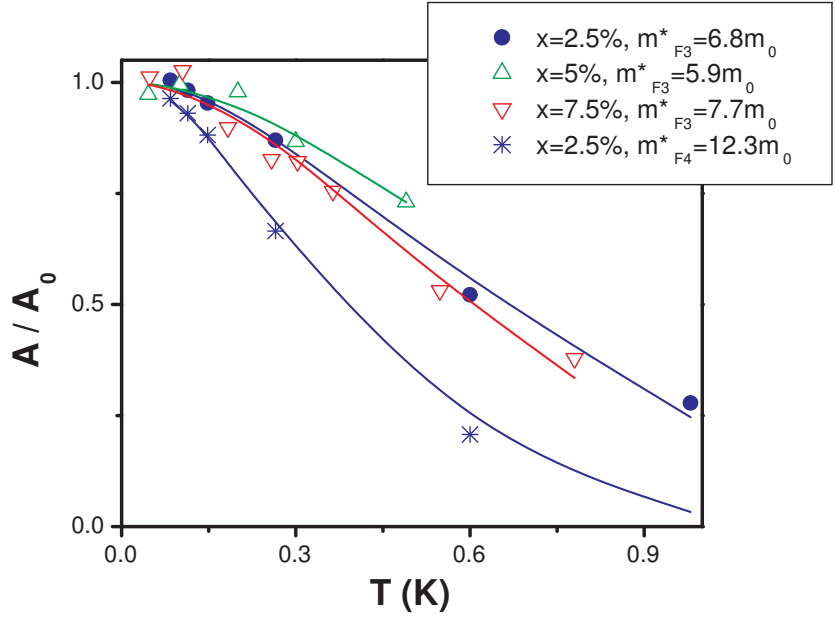}}
\caption{\label{fig:Meff} (Color online) Normalized dHvA amplitude
vs Temperature in CeCo(In$_{1-x}$Cd$_x$)$_5$. The amplitudes shown
correspond to the dHvA branches $F_3$ ($\bullet$) and $F_4$ ($\ast$)
for $x=2.5\%$, $F_3$ ($\bigtriangleup$) for $x=5\%$, and $F_3$
($\bigtriangledown$) for $x=7.5\%$. Solid lines are fits of the
Lifshitz-Kosevich formula to the data (see text).}
\end{figure}

\indent Figure~\ref{fig:Meff} shows the dHvA amplitudes as a
function of temperature for the $F_3$ and $F_4$ orbits in
CeCo(In$_{1-x}$Cd$_x$)$_5$. As the temperature is increased, the
Landau levels are broadened, and the dHvA oscillations suppressed.
This suppression is well described by the Lifshitz-Kosevich (LK)
formula\cite{Shoenberg}:$$A(T,H) = A_0 \frac{X_T}{sinh(X_T)},$$ with
$X_T=\frac{\alpha m^{\ast}T}{m_eH}$, where $A$ is the dHvA
amplitude, $A_0$ the $T=0$ amplitude, $m^{\ast}$ the effective mass,
$m_e$ the bare electron mass, and $\alpha = \frac{\pi^2 k_B}{\mu_B}
= 14.69$ T/K. The fit to the LK expression (see solid lines in
figure~\ref{fig:Meff}) allows the determination of the effective
cyclotron mass $m^{\ast}$ for each orbit. The values obtained for
$m^{\ast}$ are $6.8 \pm 0.2, 5.9 \pm 0.4$, and $7.7 \pm 0.4$ (in
units of the bare electron mass $m_0$) for the $F_3$ branch in the
$x=2.5\%,5\%$ and $7.5\%$ samples, and $m^{\ast}=12.3 \pm 0.2$ for
the $F_4$ branch in the $x=2.5\%$ sample, as listed in
Fig.~\ref{fig:Meff} and table~\ref{Table:dHvA1}. These are close to,
but smaller than, the values of $8.4$ $m_0$ and $18$ $m_0$
previously determined for $F_3$ and $F_4$ from dHvA measurements
along the same orientation ($H\parallel [001]$) in pure
CeCoIn$_5$\cite{settai2001}. The cyclotron effective masses in
LaCo(In$_{1-x}$Cd$_x$)$_5$ have not been investigated in this study
and are assumed to be comparable to the values found in pure
LaCoIn$_5$\cite{PhysRevLett.93.186405}.

\indent The most striking result of this analysis is the absence of
mass enhancement at the critical concentration $x_c=7.5\%$, at odds
with the presence of an AFM quantum critical point in the phase
diagram. Given the contrasting effect of pressure and Cd doping in
this system, and given that pressure is known to suppress $m^{\ast}$
in pure CeCoIn$_5$\cite{shishido2004}, one would naively expect Cd to
enhance $m^{\ast}$.  Similar to our results, no mass enhancement is
observed via dHvA measurements in Rh doped
CeCoIn$_5$\cite{goh:056402}, in which no change in the Fermi surface
is observed at the onset of AFM order. The absence of mass enhancement
with Cd doping may be simply due to the high magnetic fields used for
detecting dHvA oscillations and known to be detrimental to
$m^{\ast}$. We also cannot exclude a mass enhancement for fields
applied in-plane, nor for the hole sheets of the Fermi surface, as we
have only been able to determine $m^{\ast}$ for $H\parallel [001]$ on
the lightest part of the Fermi surface, namely the largest electron
sheet. The lack of mass enhancement near the QCP is similar to the
case of Cr where the QCP is not accompanied by a large carrier mass
enhancement. This has been shown to be due to the small phase space
occupied by the exchange enhanced magnetic fluctuations\cite{hayden}.

\indent In contrast, dHvA measurements on CeRhIn$_5$ under pressure
reveal a drastic change to the Fermi surface, with diverging
effective masses, at the pressure required to suppress the
antiferromagnetic state\cite{shishido2005}. In light of these
results, the absence of mass enhancement in
CeCo(In$_{1-x}$Cd$_x$)$_5$ at $x_c=7.5$\% may be related to the
absence of significant changes to the light mass Fermi surface
sheets and indicate the possibility of strong fluctuation scattering
only on specific sections of the Fermi surface. These are known as
hot spots and we may not be observing these specific FS regions in
our investigation. Such hot spots, where the cyclotron effective
mass diverges, have been previously reported in cubic
CeIn$_3$\cite{PhysRevLett.93.246401}. The large ordered moment in
CeCo(In$_{1-x}$Cd$_x$)$_5$ of $\sim 0.7 \mu_B$ per
Ce\cite{nicklas:052401} for $x>x_c$ suggests that the purported SDW
transition opens a gap over a large fraction of the FS and, further,
that the precursor fluctuations in the paramagnetic state may make
it difficult to observe the parts of the FS involved even at high
field. This may explain the absence of dHvA oscillations for much of
the FS (both hole and electron FS) for $x> 2.5$\% as large fractions
of the FS may be involved in the nesting associated with the
SDW-like state. Recently, hot spots at particular regions of the
hole FS of Ce(Rh,Co)In$_5$ have been suggested to explain the
commensurate, $Q=(1/2,1/2,1/2)$, AFM order that has been observed
between Rh concentrations of 25 and
60\%(Ref.~\onlinecite{matsuda2007}).

\begin{figure}
\resizebox{!}{0.3\textwidth}{\includegraphics{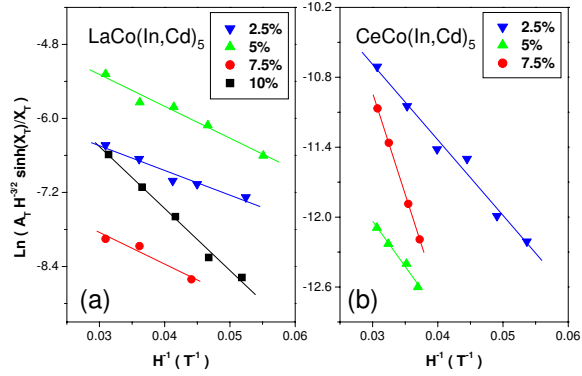}}
\caption{\label{fig:Dingle} (Color online) Dingle Plot. Reduced
amplitude (see text) vs inverse magnetic field in (a)
LaCo(In$_{1-x}$Cd$_x$)$_5$ and (b) CeCo(In$_{1-x}$Cd$_x$)$_5$ for
indicated nominal Cd concentration. Solid lines are linear fits that
determine the Dingle temperature, $T_D$.}
\end{figure}

\begin{table}
\caption{\label{Table:dHvA1}Effective mass ($m^{\ast}$), Dingle
temperature ($T_D$), mean free path ($\ell$) and residual
resistivity ratio (RRR) in LaCo(In$_{1-x}$Cd$_x$)$_5$ and
CeCo(In$_{1-x}$Cd$_x$)$_5$.}
\begin{ruledtabular}
\begin{tabular}{cccccc}
 & $x$($\%$)& $m^{\ast}$($m_0$) & $T_D$(K) & $\ell$ (nm)& RRR \\
LaCo(In$_{1-x}$Cd$_x$)$_5$&2.5& &2.71&186&58\\
 &5& &3.50&143&36\\
 &7.5& &3.48&144&26\\
 &10& &6.81&73&16\\
CeCo(In$_{1-x}$Cd$_x$)$_5$& 2.5& 6.8&0.64&119&\\
 &5&5.9&0.89&98&\\
 &7.5&7.7&1.53&43&\\
\end{tabular}
\end{ruledtabular}
\end{table}

\indent Landau levels are also broadened by impurity scattering of
the quasiparticles. In the Lifshitz-Kosevich theory, the associated
amplitude reduction factor is the so-called Dingle factor,
$exp(-\frac{2 \pi^2 k_B T_D}{\beta H})$, where $k_B$ is the
Boltzmann factor, and $\beta = g$ $\mu_B \frac{m_0}{m^{\ast}}$ with
$g$ the Land\'{e} factor. The Dingle temperature, $T_D$, is defined
as $T_D=\frac{\hbar}{2 \pi k_B \tau}$ with $\tau^{-1}$ the impurity
scattering rate\cite{Shoenberg}.  Experimentally, $T_D$ is
determined from the slope of the reduced amplitude,
$ln(\frac{A_T}{H^{3/2}} \frac{sinh(X_T)}{X_T})$, vs inverse magnetic
field, $H^{-1}$, where $A_T$ is the dHvA amplitude measured at the
lowest temperature (0.1~K for CeCo(In$_{1-x}$Cd$_x$)$_5$ and 0.3~K
for LaCo(In$_{1-x}$Cd$_x$)$_5$). The Dingle plots in both systems
are shown in Fig.~\ref{fig:Dingle}. This allows an estimation of the
mean free path defined as $\ell= v_F \tau$, where $v_F$ is the Fermi
velocity, given by $v_F=\frac{\hbar k_F}{m^{\ast}}$ with $k_F$
related to the dHvA frequency $F$ through the Onsager relation: $2e
F = \hbar k_F^2$ ($e$ being the electronic charge). We have used the
frequencies of $F_3$ shown in Fig.~\ref{fig:FreqvsCdAll} to
determine $k_F$ and the mean free path $\ell$ for this orbit for $H
\parallel [001]$ in both LaCo(In$_{1-x}$Cd$_x$)$_5$ and
CeCo(In$_{1-x}$Cd$_x$)$_5$. The results of $T_D$ and $\ell$ are
summarized in table~\ref{Table:dHvA1} together with $m^{\ast}$. The
values obtained by us are consistent with the previously
reported\cite{PhysRevLett.93.186405} $\ell \simeq 200$~nm and 70~nm
in LaCoIn$_5$ and CeCoIn$_5$ respectively. We found that in both
compounds there is a systematic suppression of the mean free path
due to disorder scattering introduced by the Cd substitution. A
similar suppression is observed in the residual resistivity ratios
(RRR)with increasing Cd concentrations for
LaCo(In$_{1-x}$Cd$_x$)$_5$ (see table~\ref{Table:dHvA1}), with the
RRR defined as the ratio of the zero field resistivity at 300 and 3
K (not shown). The RRR values in CeCo(In$_{1-x}$Cd$_x$)$_5$ do not
directly reflect the degree of disorder in the material due to the
presence of the coherence peak and therefore they have been omitted
in table~\ref{Table:dHvA1}.

\section{Discussion and Conclusions}

\indent The most obvious effects of Cd substitution into CeCoIn$_5$
and LaCoIn$_5$ that our data reveal are the systematic lattice
contraction, as we established with X-ray diffraction, and the
chemical potential shift due to the hole doping as apparent from the
analysis of our dHvA data in LaCo(In,Cd)$_5$. We demonstrated that
for LaCo(In,Cd)$_5$ the dHvA frequencies associated with the main
electron Fermi surface sheet decrease, while those of the hole sheet
increase in a manner consistent with hole doping with Cd
substitution. In addition, we demonstrated that a dHvA frequency
associated with the electron sheet of Cd doped CeCoIn$_5$, which is
the only piece of the Fermi surface resolved in our dHvA data,
decreases at a rate similar to the La analog, again consistent with
that expected for a small density of doped holes. The corresponding
change in the electron FS volume only accounts for 1/30 of the doped
hole/Cd, suggesting that the added holes are mainly distributed over
the remaining pieces of the Fermi surface, which we do not observe.
Overall, the Fermi surface of CeCo(In$_{1-x}$Cd$_x$)$_5$ remains
closely related that of pure CeCoIn$_5$, with only modest changes in
the dHvA frequencies and cyclotron masses, despite the dramatic
evolution of the zero field ground state from superconducting to
superconducting+antiferromagnetic. The similarity of the changes
that occur with doping in the La and Ce compounds allows us to rule
out any substantial $f-$electron localization and, thus, to rule out
mechanisms for the antiferromagnetic state that rely on local moment
formation. It follows that the commensurate AFM order in
CeCo(In$_{1-x}$Cd$_x$)$_5$ is likely due to an itinerant, SDW-type,
mechanism which relies on FS nesting. The most famous example of SDW
ordering that is commensurate with the underlying lattice is the
elemental antiferromagnet Cr which, although incommensurate when
pure, evolves to a commensurate state with small Mn
doping\cite{fawcett}.

\indent Our dHvA data are also similar to that of Rh doped
CeCoIn$_5$\cite{goh:056402} where the FS is seen to undergo small
changes so that a large Fermi surface is observed in both Cd and Rh
doped CeCoIn$_5$. Thus, for both Cd and Rh substitution into
CeCoIn$_5$ commensurate AFM order coexisting with superconductivity
is observed along with a Fermi surface that appears to contain a
substantial contribution from the Ce 4$f$-electrons. The main
difference in these two substitution series is that Cd substitution
suppresses superconductivity\cite{pham:056404} for concentrations
beyond 15\%, while superconductivity remains apparent up to 60\% Rh
substitution\cite{PhysRevB.65.014506}. The stronger suppression of
superconductivity with Cd may be the consequence of in-plane
impurity scattering.

\indent However, there are several aspects of Cd doped CeCoIn$_5$ that
remain poorly understood. It is well known that pressure applied to
CeCo(In$_{1-x}$Cd$_x$)$_5$ causes a return to paramagnetism and an
increase of the superconducting critical temperature so that pressure
appears to reverse the most obvious consequences of Cd
doping\cite{pham:056404}. If the main effect of Cd doping into
CeCoIn$_5$ is a shift of the chemical potential caused by the addition
of holes as our data suggest, then it is difficult to account for the
reversible tuning of the AFM order with pressure. For any reasonable
value for the compressibility of Cd doped CeCoIn$_5$ the carrier
density change with experimentally accessible pressures would be very
small. Thus, it is unlikely that pressure simply reverses the changes
that occur with Cd doping. This suggests that there are subtle changes
that occur to CeCoIn$_5$ with doping or pressure that are more likely
associated with the Kondo effect and the formation of the heavy
fermion metallic state. A second, perhaps related important open
question, and perhaps a clue to the origin of the AFM order, is why
the magnetic structure is commensurate, with the same wavevector, $Q=
(1/2,1/2,1/2)$, as the neutron scattering resonance observed in
superconducting, nominally pure, CeCoIn$_5$\cite{stock:087001}. In
addition, the lack of more direct evidence for SDW formation leaves
open the possibility that the magnetic state in
CeCo(In$_{1-x}$Cd$_x$)$_5$ has a character intermediate between local
moment or highly itinerant so that a simple description is difficult.

It appears from our data, as well as from the NMR
results\cite{urbano:146402}, that Cd doping of CeCoIn$_5$ into an
AFM phase does not conform to the Doniach model\cite{Doniach} where
the Kondo and RKKY coupling compete at a quantum critical point.
Instead our data suggest that a more itinerant antiferromagnetism
develops out of a Fermi surface which contains the hybridized Ce
4$f$-electrons.  The role of Cd for inducing this AFM order in
CeCoIn$_5$ remains elusive and the resolution of this mystery is
likely to broaden our approach to quantum criticality beyond the
Doniach phase diagram.

\begin{acknowledgments}
We are thankful to E.C. Palm for his technical assistance with the
use of dilution refrigerator. A portion of this work was performed
at the National High Magnetic Field Laboratory, which is supported
by NSF Cooperative Agreement No. DMR-0654118, by the State of
Florida, and by the DOE. L.B. is supported by DOE-BES. R.G.G. was
supported directly by NSF. C.C. acknowledges ICAM fellowship. Z.F.
acknowledges support through NSF Grant No. NSF-DMR-0503361. J. F. D.
acknowledges support through NSF Grant No. NSF-DMR-084376. D.P.Y.
acknowledges support through NSF Grant No. NSF-DMR-0449022. J.Y.C.
acknowledges support through NSF Grant No. NSF-DMR-0756281. I. V.
was supported in part by DOE Grant. No. DE-FG02-08ER46492.
\end{acknowledgments}

\appendix*

\section{}

\indent We present below an estimation of the volume change of the
electron Fermi surface in CeCo(In$_{1-x}$Cd$_x$)$_5$ due to Cd
within cylindrical Fermi surface approximation. We use the Onsager
relation:\\
\begin{equation}
  F=\frac{\hbar c}{e}A=\frac{1}{\pi}\Phi_0 A\,,
\end{equation}

where $\Phi_0=hc/2e=2\cdot 10^{-11}$T$\cdot$cm$^2$ is the flux
quantum. The shift in the frequency of the $F_3$ (electron) orbit,
$\delta F$, translates into the change in the area of the extremal
orbit, and allows for a rough estimate of the change in the volume
of the Fermi surface via:\\

\begin{equation}
  \delta V=\frac{2\pi}{l_c}\delta A=\frac{2\pi^2\delta
  F}{l_c\Phi_0}.
\end{equation}

where $l_c$ is the lattice constant along [001]. The number of
states in this volume is (with a factor of 2 for spin degeneracy):\\

\begin{equation}
  \delta n=\frac{2 \delta V}{(2\pi)^3}=\frac{1}{2\pi}\frac{\delta
  F}{l_c \Phi_0}\,.
\end{equation}

\indent Using experimental values of $\delta f\approx 2.5\cdot
10^2$T per 10\% nominal Cd, and $l_c=7.6$\AA, we get $\delta n
\approx 2.6 \times 10^{19} \mbox{cm}^{-3}$. The next step is to
determine what fraction of 1 hole per Cd this change in density
corresponds to. With a unit cell volume $v_u\approx 161
\mbox{\AA}^3=1.6\times 10^{-22}\mbox{cm}^3$ and given that each unit
cell has $5x$ holes, the density of added holes is:
$\frac{5x}{v_u}\approx 3.1 x \times 10^{22}\mbox{cm}^{-3}$. For
nominal $x=0.1$ we expect the actual Cd concentration to be $x \sim
0.03$, so we should have $\delta n\approx 9\times
10^{20}\mbox{cm}^{-3}$. In other words, the change in the electron
Fermi surface volume (estimated from the change in the $F_3$
frequency) due to Cd only accounts for $\sim 3\%$ of the additional
hole, assuming that each Cd introduces one hole.

%\bibliography{Cd115References}

\end{document}